**Comment on :** *Testing the speed of the "spooky action at a distance" in a tabletop experiment.*
(*Sci Rep* **13**, 8201 (2023). https://doi.org/10.1038/s41598-023-35280-8)

Bruno Cocciaro[a], Sandro Faetti[b] & Leone Fronzoni[c]

  a- Liceo Scientifico XXV Aprile, via Milano 2, 56025 Pontedera(Pisa) Italy. email: b.cocciaro@comeg.it
  b- retired Associate professor of the Dipartimento di Fisica dell'Università di Pisa, Largo B. Pontecorvo, 56123 Pisa, Italy. email: sandro.faetti@unipi.it.
  c- retired Associate professor of the Dipartimento di Fisica dell'Università di Pisa, Largo B. Pontecorvo, 56123 Pisa, Italy. email: leone.fronzoni@unipi.it.

**Abstract**
In 1989, Eberhard proposed a *v*-causal model[1] where quantum correlations between entangled particles are established by communications moving at a superluminal speed $v_t = c\,\beta_t$ ($\beta_t > 1$) in a preferred frame. In successive years, several experiments[2-9] established lower bounds $v_{t,max} = c\beta_{t,max}$ for the possible tachyons velocities. In a recent paper[9], Luigi Santamaria Amato et al. performed an interesting east-west aligned tabletop experiment under the assumption that the preferred frame is the Cosmic Microwave Background (*CMB*). In that paper, they criticize long-distance experiments but here we show that most of their criticisms are not applicable to long-distance tunnel experiments[7,8] where the highest lower bound was obtained.

Santamaria Amato et al. say that long distance experiments are affected by many drawbacks and, in particular: *uncontrollable environment conditions, non-perfect east-west alignment, no measure of the coherence length and days-long measurement time.* We remind that a tabletop experiment oriented along the east-west direction was already published by our group[4]. However, our main interest is to show that most of these sentences cannot be applied to the long-distance experiments[7,8] carried out in a tunnel of the European Gravitational Observatory (*EGO*). In these experiments, the temperature of the central bench where the entangled photons and the reference beams were generated was kept fixed within 0.1 °C and an efficient compensation procedure[10] was used to get a high intensity (≈1300 coincidences/s) and a high fidelity source of entangled photons with average fringes visibility higher than 94% both in the laboratory preliminary measurements and in the final tunnel measurements. This clearly shows that the tunnel environment conditions did not affect appreciably the purity of the quantum entangled state. Furthermore, thanks to suitable feedback procedures[7,8], both the trajectories of the entangled beams and the equality of their optical paths were kept fixed within a small uncertainty thus ensuring a great stability of the experimental results over the entire days-long measurement time (see, for instance, figures 9a) and 9b) in reference[7]). Then, it is evident that the environment conditions can be well controlled in long distance tunnel experiments.

The coherence length of the entangled photons was already shown to be negligible[4]. In all our experiments[4,7,8], we used two thin adjacent nonlinear *BBO* crystals (≈ 0.5 mm thickness) cut for type-I phase matching that produced entangled photons with average wavelength $\lambda_d = 813$ nm and spectral width $\Delta\lambda_d \approx 70$ nm. Direct measurements of the coherence length $L_c$ of down-converted photons produced by irradiation of a nonlinear crystal have been performed by many authors[11-13]. In particular, experiments with interference filters[12,13] (bandwidth $\Delta\lambda_F < \Delta\lambda_d$) showed that the coherence length is in a satisfactory agreement with the Gaussian expression $L_c \approx 2\ln(2)\,\lambda_d^2/(\pi\,\Delta\lambda_F)$. Substituting the experimental values $\Delta\lambda_F = 40$ nm and $\lambda_d = 813$ nm we get $L_c \approx 7.3$ μm that is completely negligible compared to the other error sources leading to the experimental uncertainty $\Delta d = 215$ μm ($\sqrt{(215)^2 + (7.3)^2} = 215.12$).

The Alice (A) and Bob (B) polarizers were not aligned along the east-west direction in the tunnel experiments[7,8] but, as noticed in[8], only a small fraction (5%) of all possible preferred frames was inaccessible (the *CMB* frame was fully accessible). Note that an east-west alignment of the Alice and Bob polarizers was planned in our initial project[6] where the two Ego tunnels should have been connected with a 30 cm-diameter pipe. Only reasons of time and budget did not allow to follow the initial project, but there should not be any problem carrying out tunnel experiments with the east-west geometry[6].

Finally, Santamaria Amato et al. criticized long distance measurements and "*days-long measurement time*" and say that their measurements allow to "*avoiding of the Bell test splitting in several days*". Note that any complete measurement always requires at the least a 12 hours-measurement time[1]. To answer to the criticism above we need to delve a little into the basic aspects of the standard measurement of the lower bound

$$\beta_{t,max} = \sqrt{1 + \frac{(1-\beta^2)(1-\rho^2)}{\left[\rho + \frac{\omega \beta \sin \chi \; \delta t}{2}\right]^2}} \; , \tag{1}$$

where $\rho = \Delta d/d$, $\Delta d$ is the uncertainty on the equalization of the photons optical paths, $d$ is the distance between the Alice and Bob polarizers, $\delta t$ is the time that is needed to perform a measurement of the Bell parameter $S$ and $\omega = 7.29 \times 10^{-5}$ rad/s is the Earth angular velocity. Parameters $\beta$ and $\chi$ characterize modulus and orientation of the unknown reduced velocity $\boldsymbol{\beta} = \boldsymbol{v}/c$ of the preferred frame. The purpose of any experiment should be to establish the highest values of $\beta_{t,max}$ for any preferred frame ($\beta < 1$ and $\sin \chi \leq 1$). According to eq.(1), this goal requires $\rho = \Delta d/d \ll 1$ that is greatly favoured by long-distance experiments ($d > 1$ km) where it has been possible to get[8] $\rho = 1.83 \times 10^{-7}$ ($\rho \geq 2.6 \times 10^{-5}$ in the tabletop experiments[4,9]). Furthermore, velocity $\beta$ of the preferred frame is unknown and, thus, $\beta_{t,max}$ should be maximized for any possible preferred frame ($\beta < 1$ and $\sin \chi \leq 1$). This occurs if $\delta t \ll 2 \rho/\omega$ where $\beta_{t,max}$ in eq.(1) reduces to $\beta_{t,max} \approx \sqrt{(1-\beta^2)}/\rho$ that is virtually independent of $\beta$ for not relativistic reference frames ($\beta^2 \ll 1$). The standard method to measure $\beta_{t,max}$ requires at the least 8 successive measurements of coincidences with different orientations of the polarizers to obtain the Bell parameter $S$. Unfortunately, the time wasted rotating the polarizers leads to somewhat long acquisition times ($\delta t \gg 2 \rho/\omega$) as occurred in the case of our preliminary measurement[7] in the *EGO* tunnel where we obtained $\delta t = 200$ s (100 s in[9]) leading to the green curve in fig.1. Time $\delta t \gg 2 \rho/\omega$ is the cause of the reduction of $\beta_{t,max}$ for $\beta > 10^{-5}$. For this reason, a completely different but rigorous procedure ("*the Bell test splitting in several days*") needing measurements made in at least 4 days was proposed (see page 052124-8 in reference[8] for details) that allowed to reduce $\delta t$ up to $\delta t = 0.492$ s leading to the red curve in fig.1 with $\beta_{t,max} \approx 5 \times 10^6$ for the *CMB* frame ($\beta \approx 1.3 \times 10^{-3}$, $\chi = 83.6°$). In fig.1 we also show, for comparison, the blue curve that would be obtained substituting $\sin \chi = 1$ and the Santamaria et al.[9] parameters $\rho$ and $\delta t$ into eq.(1). The advantage of using both the long-distance experiments and the new procedure is evident in figure 1 and, thus, we infer that this procedure represents an improvement of the standard method.

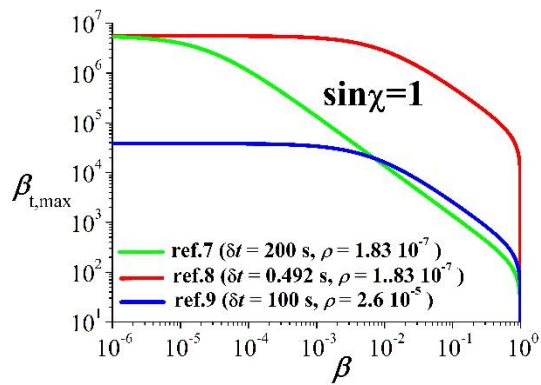

**Figure 1**: Lower bound $\beta_{t,max} = v_{t,max}/c$ versus the dimensionless velocity $\beta = v/c$ of the preferred frame.

Of course, this procedure cannot be used to close the locality loophole because the polarizers orientations are changed only at the beginning of each measurement day. However, we note that almost all experiments[2-4,6-9] concerning the *v*-causal models, including that of Santamaria Amato et al., did not take the locality loophole into account. At the best of our knowledge, only Yin at al.[5] measured the speed of "spooky action at a distance" on long distance experiments without locality and measurement choice loopholes showing that these loopholes are not responsible for quantum correlations, in agreement with the results of a large number of other Bell experiments. Then, we agree here with the sentence of Salart et al.[14]: "*we consider this point as established and we address another issue: the alternative assumption that quantum correlations are due to supra-luminal influences of a first event onto a second event*".